# Integrated Key based Strict Friendliness Verification of Neighbors in MANET

Dhadesugoor R. Vaman*

Department of Electrical and Computer Engineering
Prairie View A & M University
Prairie View, Texas, USA
e-mail: drvaman@pvamu.edu
* TI Endowed Chair Professor and Director of ARO CeBCom

Niraj Shakhakarmi**

Department of Electrical and Computer Engineering
Prairie View A & M University
Prairie View, Texas, USA
e-mail: nshakhakarmi @pvamu.edu
** Doctoral Researcher

*Abstract*—A novel Strict Friendliness Verification (SFV) scheme based on the integrated key consisting of symmetric node identity, geographic location and round trip response time between the sender and the receiver radio in MANET is proposed. This key is dynamically updated for encryption and decryption of each packet to resolve Wormhole attack and Sybil attack. Additionally, it meets the minimal key lengths required for symmetric ciphers to provide adequate commercial security. Furthermore, the foe or unfriendly node detection is found significantly increasing with the lower number of symmetric IDs. This paper presents the simulation demonstrating the performance of SFV in terms of dynamic range using directional antenna on radios (or nodes), and the performance in terms of aggregate throughput, average end to end delay and packet delivered ratio.

*Keywords- Integrated key; Strict Friendliness Verification; Mobile Ad hoc Networks ; Wormhole Attack; Sybil Attack; Foe Detection Rate.*

I. INTRODUCTION

Geographic based friendly node identification requires location tracking and moving patterns of the radios within MANET environment. Using directional antenna on each node can provide better spatial use of bandwidth and energy efficient preventive solutions for wormhole attacks. Additionally, symmetric IDs can be dynamically changed to solve Sybil attack of multiples false identities. All these issues are required to be addressed for strict friendliness verification between neighbor nodes in MANET, before nodes participate in the direct or multi-hop communications. This is accomplished by packets encryption and decryption using robust integration of different partial keys for each packet so that neighbors can be declared as strict friendly radios (also referred to as nodes). For encryption and decryption, a set of partial keys are generated from the private information of nodes that includes location information (including distance and direction of node), symmetric ID and round trip time (RTT) of preamble packet between two neighboring nodes to maintain the anonymity. This prevents possible replay from the Wormhole attackers, Sybil attackers and foe nodes.

This paper is organized as follows: Section II provides the related work; Section III describes anonymous geographic parameters; Section IV describes security against Wormhole attacks and Sybil attacks; Section V describes proposed Strict Friendliness Verification scheme; Section VI derives Foe (Unfriendly) nodes detection rate; Section VII describes performance & evaluation and followed by conclusion in Section VIII.

II. RELATED WORK

Wormhole prevention mechanism deploys 'packet leashes' containing timing and Global Positioning System (GPS) information to each packet on a hop-by-hop basis [1] and end to end basis [2] to verify the actual physical distance covered by packets. Furthermore, an end-to-end mechanism based on geographic information detects anomalies in neighbor relations and node movements [3]. But, it has some drawbacks that it misses some anomalies as the records falling into the same slot and the same cluster might be ignored. Additionally, detection of malicious nodes is to protect the location discovery and detection of replay signals is to avoid false positives services [4]. These issues can be addressed by deploying the proposed integrated key based security scheme.

A directional information sharing can prevent wormhole endpoints from being camouflaged as false neighbors and reduce the intimidation of wormhole attacks, without any location information or clock synchronization [5], [7]. This scheme does not address the prevention from multiple endpoint attacks, which requires substantial amount of energy consumption. On the other hand, secure localization (SeRLoc) uses the geometric and radio range information to detect the wormhole attack and the Sybil attack on localization scheme in which only few nodes are equipped with directional antennas [8] which is an issue in the overall design of a MANET. These can be alleviated by the proposed scheme by using location information, symmetric IDs and RTT assets to generate integrated key for strict friendly verification and multiple endpoints attacks cannot succeed any more.

Related key cryptanalysis is a real-time attack in the key integrity on the key-exchange protocols by an attacker flipping bits in the key without knowing the key and key function. This can be resolved by designing keys such that a key in each round is randomly generated to prevent any related key cryptanalysis [12]. This can be made more advanced by using different partial keys generated from random number generator functions on each round keys using different seeds so that related key cryptanalysis can be deciphered. On the other hand, it is necessary to increase the key sizes gradually for effective countermeasures against new cryptanalytic insights to maintain a comfortable margin of security. This is based on explicitly formulated parameters and existing cryptosystems in symmetric cryptosystems, RSA, and discrete logarithm based cryptosystems [13]. This key size issue is addressed by designing optimal bits length integrated key scheme for commercial security.

### III. ANONYMOUS GEOGRAPHIC PARAMETERS

Each radio or node has unique geographic location and RTT assets which can be deployed as anonymous geographic parameters for strict friendliness. The location information includes radial or scalar Euclidean distance and angle between the sender and receiver which are computed from the average time difference of TOA (Time of Arrival) and TOD (Time of Departure) and the angular bearing by AOA (Angle of Arrival) for $i^{th}$ number of packets. The radial distance ($D_{Radial}$) between sender and receiver is computed as the average time difference of arrival time of preamble packet at receiver and departure time of response from receiver to sender as follows:.

$$D_{Radial} = c[\frac{1}{n}\{\sum_{i=1}^{n}(ToD_i - ToA_i)\}] \quad (1)$$

$$RTT = T_1 - T_2 \quad (2)$$

where,

Speed of light (c) = $3\times10^8$ m/ sec
No. of packets for ranging (n) = 3 packets
Transmission time from sender to receiver ($T_1$)
Responding time from receiver to sender ($T_2$)

In AOA, directional antenna-arrays are used to estimate the direction of arrival (θ) of the signal of interest and a single AOA measurement constrains the source along a line. The precision of AOA depends upon the line of sight between sender and receiver. Additionally, RTT (Round Trip Time) is computed as the total time elapsed between the transmission and the reception of the acknowledgement for $i^{th}$ number of packet. This round trip time includes the propagation delay of the packet traveling in both directions and processing delay. The computed $D_{Radial}$, $AOA$ and $RTT$ assets must be lower or equal to their corresponding maximum range value, $D_{Max\_Radial}$, $AOA_{Max}$ and $RTT_{Max}$ so that the sender's request has not been "replayed" by wormhole or local nodes and proceed as seeds towards strict friendliness. Otherwise, the sender's request is rejected and needs to repeat with a second attempt with another $D_{Radial}$, $AOA$ and $RTT$ assets.

### IV. SECURITY AGAINST WORM HOLE AND SYBIL ATTACKS

Wormhole attack is the direct network link to eaves drop messages at one point of the network and replay at another point, which sever multi hop spatial reuse in mobile wireless ad hoc networks. Wormhole attacks deploy the encrypted packets which they overhear from the legitimate nodes and replay them to create a major issue in filtering those packets by any preventive cryptographic measures. In Fig. 1, node A has off channel link known as tunnel to node B and replay cipher packets between node C and D. This issue can be addressed by using real time location information achieved from directional antenna, symmetric ID and RTT assets to generate the integrated key for strict friendly verification before multi hop routing of packets. Real time geographical information and RTT cannot be replayed by the wormhole attackers as they are only virtual nodes with off-channel link. Even though they become capable to replay the encrypted packet with the integrated key, it will be no longer accepted by strict friendly verifier because two identical encrypted packets do not exist in the proposed SFV scheme. The reasoning is that the integrated key is dynamically generated from the pseudo random number generation functions using two different seeds (Location, RTT) as well as data packet, and changeable symmetric ID.

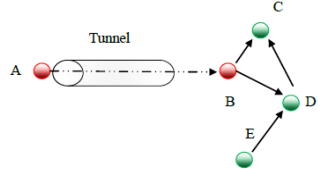
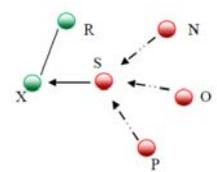

Fig. 1 Wormhole Attack    Fig. 2 Sybil Attack

Sybil attack is the adversary or illegitimate use of the multiple identities to function as distinct nodes masquerading. It is similar to wormhole attack as it also replays the encrypted packets but using false identity. This can attack the distributed storage, routing, data aggregation, voting, fair resource allocation and misbehavior detection. In Fig. 2, node S adversary uses the secret keys (identities) of three nodes N, O and P to replay node X as three different nodes. Node X gets the false realization of three more direct neighbors N, O and P except node R and S. This type of bogus polymorphism identities are commended to be strictly verified by a legitimate node. This issue is addressed by using a bunch of symmetric IDs along the location information and RTT for the integrated key generation. The symmetric ID of a node is dynamically changed on each link establishment to another node. Additionally, each packet is encrypted or decrypted by different integrated key and replay cannot fraud verifier, which possesses its robustness.

## V. PROPOSED STRICT FRIENDLINESS VERIFICATION SCHEME

Neighbor nodes are essential to legitimately verify them as strictly friends so that they can co-operate in the secured location tracking through multi-hop communication. The novel key in encryption and decryption for the strict friendliness verification deploy 90 bits length integrated key and 12 bytes block size, which accomplish the minimal key lengths for symmetric ciphers to provide adequate commercial security. The novel integrated key is generated as 90 bits key K= ($K_1$ $K_2$ $K_3$) consisting 32 bits key ($K_1$), 26 bits key ($K_2$) and 32 bits key ($K_3$) to encrypt the packet at the transmitter end. Each node generates $K_1$ as pseudo random number using initial seed (i-1) in the $RNG_1$ function, $K_2$ is 26 bits unique $ID_{Tx}$ of a node and $K_3$ is generated by the encryption of the first packet with the random number generated from $RNG_2$ function using initial seed (n-1). Furthermore, the initial seed (i-1) for $K_1$, is the location information (distance and direction) between transmitter and receiver whereas the initial seed (n-1) for $K_2$, is the RTT of preamble packet between them. Then, first ensemble packet is encrypted using the first key K generated from the integration of the location information, RTT and symmetric ID. Similarly, the second key K´= (K´$_1$ K´$_2$ K´$_3$) is the integration of K´$_1$, generated from $RNG_1$ function using first halve of key K as seed i, K´$_2$ same as $K_2$ (symmetric ID) and K´$_3$, generated from $RNG_2$ function using second halve of key K as seed n. Then, the second ensemble packet is encrypted using second key K´ and this encryption procedure is iterated for next packets. The cipher packets $C_j$ for $j^{th}$ number of plain ensemble packets are computed using key $k_j$ in encryption as shown in Fig. 3 using the following equations:

$$K_{i1} = RNG_1(seed_i) \quad (3)$$
$$K_{i2} = ID_{Tx} \quad (4)$$
$$K_{i3} = \{RNG_2(Seed_n) \oplus P_i\} \quad (5)$$
$$K_i = \{K_1\ K_2\ K_3\} \quad (6)$$
$$C_i = \{E_{K_i}(P_i)\} = P_i \oplus K_i \quad (7)$$

At the receiver end, the first packet is decrypted by using key K=($K_1$ $K_2$ $K_3$) where, $K_1$ is the 32 bits key generated using initial seed (i-1) which is the location information between transmitter and receiver in $RNG_1$, $K_2$ is the 26 bits unique $ID_{Rx}$ of node and $K_3$ is 32 bits key generated by the decryption of the first encrypted packets with the random number generated from $RNG_2$ function using the initial seed (n-1) which is the RTT between transmitter and receiver. Regarding second packet, the decryption is done by changing K´$_1$ and K´$_3$ generated by RNG functions with new seeds achieved after halving the previous key and this decryption procedure is iterated for next packets. The plain ensemble packets $P_j$ for $j^{th}$ number of cipher packets are computed using keys $k_j$ in decryption as shown in Fig. 4 using the following equations:

$$K_{j1} = RNG_1(seed_i) \quad (8)$$
$$K_{j2} = ID_{Rx} \quad (9)$$
$$K_{j3} = \{RNG_2(Seed_n) \oplus C_j\} \quad (10)$$
$$Kj = \{K_1\ K_2\ K_3\} \quad (11)$$
$$Pj = \{D_{K_j}(C_j)\} = C_j \oplus Kj \quad (12)$$

The receiver is declared as a friendly neighbor node, when encrypted packets from the transmitter are successfully decrypted by the receiver. It can be used for direct or multi hop communication. If not, the node is deemed as a suspicious node and is not authorized for communications. This process of strict friendliness verification between neighbors is required each time for secure path connectivity for information exchange.

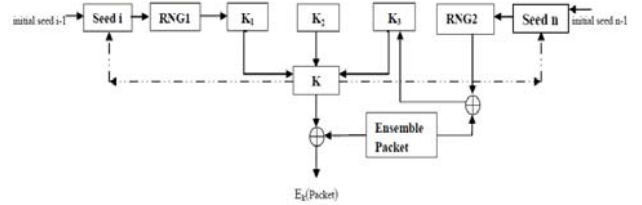

Fig. 3 Packet Encryption for Strict Friendly Verification

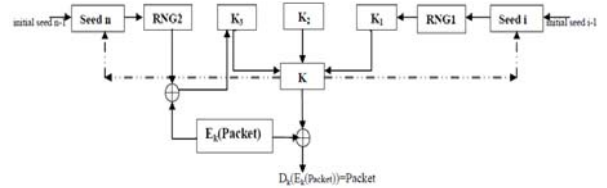

Fig. 4 Packet Decryption for Strict Friendly Verification

Regarding the symmetric cipher, higher keys length protect against brute force attacks. Increasing each bit in the key increases twice the number of possible keys and yields two times more search for the brute force attack. With 90-bits key, the complexity analysis of algorithm needs O ($2^{90}$) = 1.237940039 × $10^{27}$ runs for the brutal force search. On average, a brute force attack must check half of the total runs, performing $2^{89}$ encryptions, to find the key. This 90 bits key length is enough for the symmetric ciphers to provide plentiful commercial security.

The detection rate is the probability of detecting an unfriendly or foe node in MANET cluster which evaluates the detection performance of strict friendliness verification scheme. Let us consider a suspicious transmitting node sends request to the receiver node, such that it could persuade as a friendly node and avoid being detected during the strict friendly verification process, with the probability of replayed by wormholes ($p_{wh}$), probability of node's ID replay ($p_i$) and the probability of locally RTT replayed ($p_r$) by neighbors. Then, the probability of detection of replayed by wormholes is ($1-p_{wh}$), the probability of detection of node's ID replay is ($1-p_i$) and the probability of detection of locally RTT replayed ($1-p_r$) in the strict friendliness verification. The probability of suspicious node detection by a friendly node is computed as:

$$P = (1 - p_{wh})(1 - p_i)(1 - p_r) \qquad (13)$$

When each detecting node having n detection IDs, the detection rate or probability ($P_{dr}$) of a suspicious node being detected by a kind friendly detecting node can be computed as:

$$P_{dr} = 1 - (1 - p^n) = 1 - \{1 - (1 - p_{wh})(1 - p_i)(1 - p_r)\}^n \qquad (14)$$

This implies that unless the detection probability increases, the detection rate cannot be increased. A kind friendly detecting node can significantly increase the detection rate using higher number of symmetric IDs. Figure 5 shows that the transmitter detection rate increases with the higher number of Symmetric IDs changing at receiver because it increases the robustness against replay attacks.

## VI. SIMULATION AND PERFORMANCE AND EVALUATION

The simulation is performed using 80 nodes in each cluster of size 300 X 300 sq. m, for 10 different clusters in 3000 X 3000 sq. m terrain area deploying the proposed Strict Friendly Verification scheme for neighbor nodes as shown in Fig. 6 using simulation parameters from Table 1 [3]. The location information is computed from the average time difference of TOA and TOD of the preamble packet and the AOA of preamble packet determined by the directional antenna arrays. Similarly, RTT is computed from the TOA packets in the time domain. Both the location information and the RTT are taken using the ranging and non-ranging methods. In the ranging method, the transmission power is increased while scanning for the neighbor nodes. The first scanning is done by transmitter at the range of 230 m using steerable directional antenna and if it could not find any desired node then second scanning is done at the range of 250m by increasing the transmission power and similarly, third scanning is done at the range of 270 m. The major advantage of the ranging method is that it utilizes the minimum transmission power to the optimum extent, where as the non-ranging method uses the full power to cover the full transmission range of 270 m at the first instance.

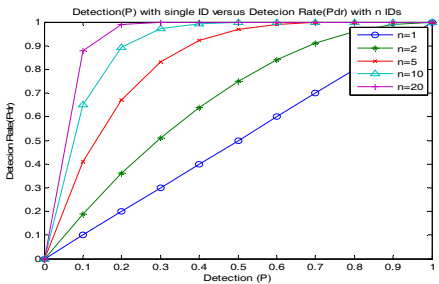

Fig. 5 Detection Rate analysis

TABLE -I
SFV SIMULATION PARAMETERS

| Simulation parameters | Values |
|---|---|
| Area (sq.m) | 3000 X 3000 |
| Radio range (m) | 230, 250, 270 |
| Nodes in each cluster | 80 |
| Transmission rate (Kbps) | 200-2000 |
| Mobility model | Random Waypoint Model |
| Traffic Type | CBR (UDP) |
| Packet size (Bytes) | 512 |
| Node speed (m/s) | 5-50 |

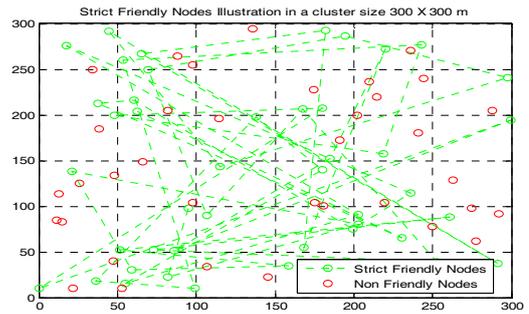

Fig. 6 Strict Friendly Verification Deployment Scenario

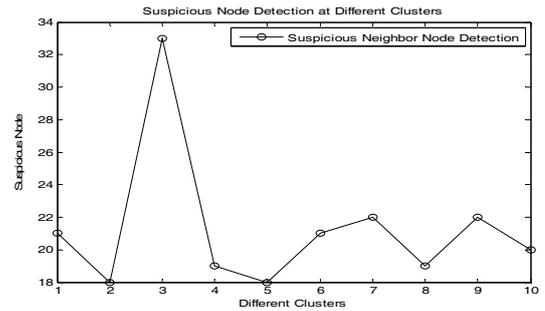

Fig. 7 Unfriendly Nodes' Detection in Different Clusters

To detect the unfriendly nodes, the transmitter generates a 90 bit key, $K = (K_1\ K_2\ K_3)$ consisting 32 bit key ($K_1$) generated from RNG1 using location information as the seed, a unique ID for the node as a 26 bit key ($K_2$) and a 32 bit key ($K_3$) generated by the encryption of the first packet with RNG2 value with RTT as the seed. This key, K is used to encrypt the first packet at the transmitter end. Similarly, the second packet is encrypted by another 90 bit key, $K'=(K'_1\ K'_2\ K'_3)$ where $K'_1$ is generated using a new seed generated from the first half of the key, K and K'$_3$ as the second half of the key, K. Similarly, receiver performs the reverse operation of above to decrypt both first and second packets using corresponding seeds and its symmetric ID. If the packets are successfully decrypted and verified, the receiving node is verified as a strict friendly neighbor, otherwise, it is deemed as a suspicious node. From our simulation, a maximum of 33 nodes are detected in third clusters as suspicious nodes as shown in Fig. 7. On the other hand, a maximum 62 are verified and declared as strict friendly neighbor nodes in the second and fifth clusters as shown in Fig. 8.

Similarly, we verified nodes as unfriendly or suspicious nodes by deploying different number of symmetric IDs during the verification process and included wormhole and RTT. Noting that the wormhole and RTT replay are external or out-of-hand to a detecting node and application of symmetric keys is in-hand for robust friendly node verification, we demonstrated the detection rate to be 30-40% for a single ID, 50 - 60 % for two IDs, 70 - 85 % for

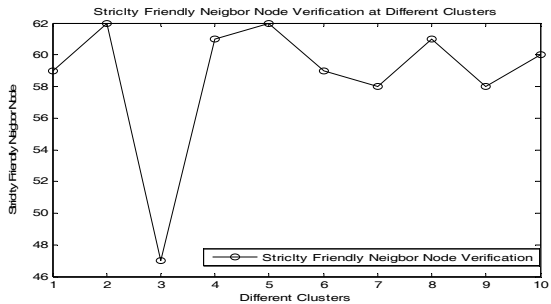
Fig. 8 Verified Strict Friendly Nodes

four IDs, 95% for six IDs and almost 100% for eight IDs. Figure 9 shows the verified strict friendly nodes in different clusters. This outstanding detection rate of almost 100% is achieved by using eight different symmetric keys for strictly friend verification in all clusters because, the ID replay entirely solved as the best case. This can be better illustrated by the detection rate analysis which shows that if the detection rate by a single ID is about 30-40% then detection rate will be approximately 100% using more than six symmetric IDs. Therefore, it can be concluded that the symmetric IDs of more than six yield outstanding detection rate, $P_{dr6} = 1-(1-P)^6$, where as the detection rate using a single ID, $P_{dr1} = 1-(1-P)$ yields about 40% resulting from $P = (1-p_{wh})(1-p_i)(1-p_r)$.

Throughput is the average rate of successful packets delivery over a communication channel which becomes saturated at a certain transmission rate when the channel capacity is fully utilized. The throughput is drastically increasing with increasing transmission rate and then saturated from the transmission rate of 600 Kbps and the saturated throughputs are approximately 1000 Kbps without SFV, 950 Kbps in SFV without ranging and 900 kbps in SFV with ranging as shown in Fig. 10. The saturated throughput is achieved when the queue scheduling optimize the successful packets' transmission.

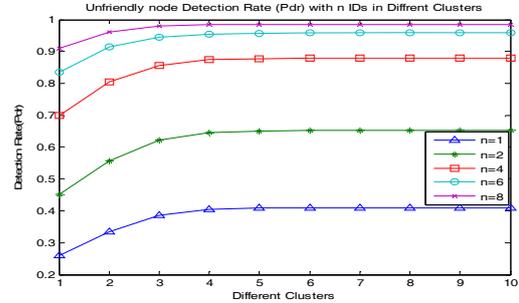
Fig. 9 Unfriendly nodes Detection Rate

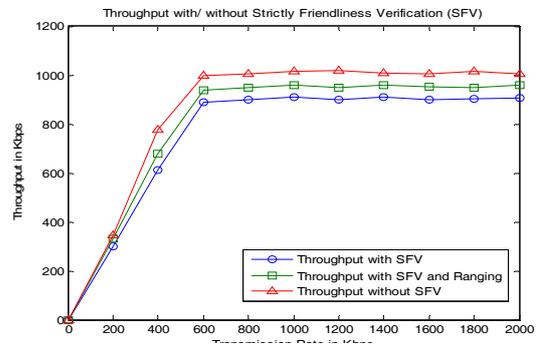
Fig. 10 Aggregate Throughput versus Transmission Rate

The average end to end delay is the time consumed for transmission of packets from the application layer of the transmitter to the reception of those packets at the application layer of the receiver. The network packets increase when the size of the queue increases and the yield increased average end-to-end delay for the delivered packets. The delay is found drastically increasing in the beginning as the transmission rate is increased up to 1000 Kbps and then saturates in all cases. The average end to end delay is found approximately saturated to 1.7 seconds using SFV, 2.0 seconds using SFV as well as ranging and 1.5 seconds without SFV as shown in Fig. 11. The saturation delay is achieved when the queue is full.

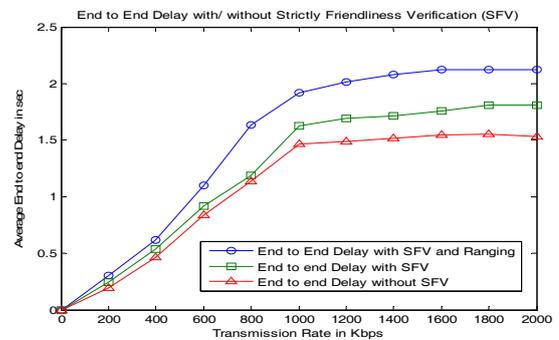
Fig. 11 Average End to end delay versus Transmission Rate

Packet delivery ratio is the ratio of the packets successfully received at receiver to packets generated at the transmitter. Packet delivery ratio decreases with increasing speed because of the link failure issue in frequently changing directional range. This result in packet delivery ratio drastically bogged down for higher mobility which is almost same up to 20 m/s in all cases and sharply drops down in SFV with ranging due to selection of range and processing delay as shown in Fig. 12.

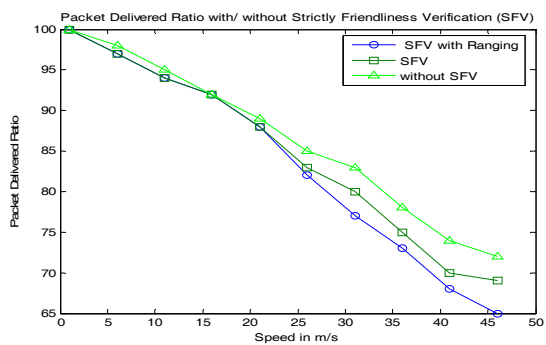

Fig. 12 Packet Delivered Ratio versus Speed

## VII. SUMMARY AND CONCLUSION

In the proposed integrated key based SFV addressed both the geographic information as well as anonymous symmetric identity of node, to resolve Wormhole and Sybil attacks in MANET. The detection rate of foe nodes is found to be 95% using six symmetric IDs in simulation. SFV with ranging have similar performance in terms of aggregate throughput, average end to end delay and packet delivered ratio as compared to SFV and without SFV. In conclusion, the SFV with dynamic ranging has significantly lower computational overhead, which makes it pragmatic and reliable in real-time co-operative MANET.

### ACKNOWLEDGMENT

This research work is supported in part by the U.S. Army Research Laboratory under Cooperative Agreement W911NF-04-2-0054 and the National Science Foundation NSF/HRD 0931679. The views and conclusions contained in this document are those of the authors and should not be interpreted as representing the official policies, either expressed or implied, of the Army Research Laboratory or the National Science Foundation or the U. S. Government.